\documentclass[%
preprint,
 amsmath,amssymb,
 aps,
]{revtex4-2}

\usepackage{graphicx}
\usepackage{dcolumn}
\usepackage{bm}
\usepackage{xcolor}
\usepackage[normalem]{ulem}

\begin{document}


\title{``Goldfish'' equations for infinitely many particles}

\author{F.~Leyvraz}
\altaffiliation[Also at ]{Centro Internacional de Ciencias, 62210, Cuernavaca, Mexico}

\email{leyvraz@icf.unam.mx}
\affiliation{
Instituto de Ciencias F\'\i{}sicas, Universidad Nacional Aut\'onoma de M\'exico 
 }%

\date{\today}
\begin{abstract}
The ``goldfish'' equations, so named because of their striking beauty, are a system of $N$ 
nonlinear ODE's, where $N$ is an arbitrary integer. They can be solved exactly in a very simple manner, by 
transforming them to free motion using a transformation involving the transition from the 
set of {\em coefficients\/} of a polynomial to that of its {\em zeroes}. This paper aims to 
explore the possibility of extending this solution to the case in which $N$ is infinite. The main difficulty involves the 
transition from polynomials to entire functions. 
\end{abstract}

\maketitle

\section{Introduction}
\label{sec:intro}

The ``goldfish'' equations, initially found in \cite{goldfish0} and further discussed
in \cite{goldfish1} where their name was coined, are the following remarkable set
of $N$ coupled non-linear ordinary differential equations:
\begin{equation}
\ddot z_j=2\dot z_j\sum^N_{k=1}{}^\prime\frac{\dot z_k}{z_j-z_k}
\label{eq:1}
\end{equation}
where the prime indicates that the sum does not include the term $k=j$. The $z_j$ are most naturally 
viewed as complex numbers and the motion therefore as a two-dimensional dynamics.
These equations are in a sense ``Newtonian'', since they express the particles' accelerations in terms of their 
positions and velocities. They obviously are not of the usual conservative type, since the forces are velocity-dependent.
However, as has been discussed in \cite{hamil,nucci,cons-laws}, they are derived from a Hamiltonian function.

These equations have the remarkable property of being exactly solvable through the following device. 
Take a monic polynomials $p(z)$ of degree $N$ and a polynomial $q(z)$ of degree $N-1$. Then, define the polynomial
\begin{equation}
\Phi(z,t)=p(z)+tq(z).
\label{eq:defPhi}
\end{equation}
This has the consequence that $\Phi(z,t)$ is monic for all $t$
.One then finds that the condition
\begin{equation}
\Phi\left[
z(t),t
\right]
=0
\label{eq:2}
\end{equation}
describes $N$ functions $z_k(t)$ for $1\leq k\leq N$, since in the complex plane every polynomial 
of degree $N$ has $N$ zeroes. (\ref{eq:1}) is found to be equivalent to the condition
\begin{equation}
\frac{d^2}{dt^2}\prod_{k=1}^N\left[
z-z_k(t)
\right]=0. 
\label{eq:2bis}
\end{equation}
so that the solutions of (\ref{eq:2}) indeed always satisfy (\ref{eq:1}).

Furthermore, any set of initial conditions
$z_j(0)$ and $\dot{z}_j(0)$ can be obtained through the following choice of the polynomials $p(z)$ and $q(z)$:
\begin{subequations}
\begin{eqnarray}
p(z)&=&\prod_{k=1}^N\left[z-z_k(0)\right]
\label{eq:3a}
\\
q(z)&=&-\sum_{j=1}^N\dot{z}_j(0)\prod_{k=1(k\neq j)}^N\left[z-z_k(0)\right]
\label{eq:3b}
\end{eqnarray}
\label{eq:3}
\end{subequations}
The system (\ref{eq:1}) can thus be reduced to free motion by transforming from the $z_k$ to the coefficients
of the polynomials $p(z)$ and $q(z)$ which are given by the symmetric functions of the $z_k$. 
{   Since the transformation from zeroes to coefficients is a point transformation} and
since free motion
is Hamiltonian, the system (\ref{eq:1}) must be so as well. 

The correspondence between the system (\ref{eq:1}) and its solution (\ref{eq:2}) also allows to obtain qualitative results, such
as the existence of unique solutions of (\ref{eq:1}) for all times for almost all initial conditions, the exceptions being such
conditions as yield, at a given finite time, a polynomial with multiple zeroes. To avoid this singular behavior, we 
allow time to vary on a complex contour going from zero to infinity, which may be so chosen as to avoid these
singularities.

%
Finally, the very striking nature of the results obtained, and the fact that the dynamics in the complex plane 
appears tantalizingly close to permitting a mechanical interpretation, have led to the study
of many variations upon it \cite{goldfish-var1,goldfish-var2,book}. 

The purpose of this paper is to extend this to the case of infinitely many $z_j$. Letting $N\to\infty$
in the original system of equations is not feasible. Rather, we shall attempt to generalize the mode of solution. 
To this end, we must 
first modify the normalization of the polynomials, and consider polynomials $p(z)$ with $p(0)=1$ and $q(z)$
with $q(0)=0$, so that the zeroes $z_j$ of $p(z)$ satisfy
\begin{equation}
p(z)=\prod_{k=1}^N\left(1-\frac{z}{z_k}\right).
\label{eq:norm}
\end{equation}
We now consider the zeroes $z_k(t)$ of $\Phi(z,t)=p(z)+tq(z)$
With this new normalization, (\ref{eq:1}) changes to
\begin{equation}
-z_j(t)\frac{d^2}{dt^2}\left[z_j(t)\right]^{-1}=\frac{\ddot{z}_j}{z_j}-2\,\frac{\dot{z}_j^2}{z_j^2}
=2\dot z_j\sum^N_{k=1}{}^\prime\frac{\dot z_k}{z_k(z_j-z_k)},
\label{eq:1bis}
\end{equation}
{    which is simply (\ref{eq:1}) where $z_j$ is replaced by $z_j^{-1}$}
The correspondence between the initial conditions
and the polynomials $p(z)$ and $q(z)$ is then as follows:
\begin{subequations}
\begin{eqnarray}
p(z)&=&\prod_{m=1}^N \left(
1-\frac{z}{z_m(0)}
\right),
\label{eq:new1a}\\
q(z)&=&z\sum_{l=1}^N\frac{\dot{z}_l(0)}{z_l(0)^2}\prod_{m(\neq l)}\left(
1-\frac{z}{z_m(0)}
\right).
\label{eq:new1b}
\end{eqnarray}
\label{eq:new1}
\end{subequations}
We thus have $p(0)=1$ and $q(0)=0$, both of degree $N$. {    This system taken in the limit $N\to\infty$
will be the central object of this paper. We discuss the details of the solution of the {\em finite\/} system in the following Section}.

\section{Basic features of the modified goldfish equations}

Let us sketch the derivation {    of the solution presented in Section \ref{sec:intro}}, where we shall emphasize some points which, while trivial in 
the polynomial case, become relevant in the case of entire functions.
Since we assume that the degree of $q(z)$ is equal to that of $p(z)$, the sum $\Phi(z,t)=p(z)+tq(z)$
is a polynomial of degree $N$ satisfying $\Phi(0,t)=1$ for all $t$, and hence has exactly $N$ zeroes.

It is clear that $z_m(0)$ must be a zero of $p(z)$, so that (\ref{eq:new1a}) immediately follows
from the assumed normalization of $p(z)$. Taking the derivative of (\ref{eq:2}) at $t=0$ yields
\begin{equation}
p^\prime\left[
z_m(0)
\right]\dot{z}_m(0)+q\left[
z_m(0)
\right]=0. 
\label{eq:defq}
\end{equation}
From this follows that we know the values of $q(z)$ at the $N$ points $z_m(0)$ as well as at $z=0$, 
which allows to compute the only polynomial of degree $N$ having this property. Specifically
\begin{equation}
q(z)=z\sum_{l=1}^N\frac{\dot{z}_l(0)}{z_l(0)^2}\prod_{m(\neq l)}\left(
1-\frac{z}{z_m(0)}
\right)
,
\label{eq:resq}
\end{equation}
which corresponds to the well-known result for Lagrange interpolation.

Let us point out an important fact at this stage: it is important to assume that $p(z)$ and $q(z)$
have {\em the same degree}. 
If $p(z)$ and $q(z)$ have different degrees, specifically
when $q(z)$ has degree $M$ larger than $N$, the degree of $p(z)$, a singular behavior ensues. 
In that case, the $z_m(t)$, which are now defined
for $1\leq m\leq M$, have a singularity near $t=0$: indeed, $N$ among the $z_m(t)$ converge to the zeroes of $p(z)$ as $t\to0$,
whereas the remaining $M-N$ values diverge as $t^{-1/(M-N)}$. 

%
Another important point concerns the determination of $q(z)$ from the initial values $z_m(0)$ and $\dot{z}_m(0)$.
As we see from (\ref{eq:defq}), the determination involves the solution of an interpolation problem. 
As stated above, for the polynomial case, if $q(z)$ has degree $N$, the solution of this interpolation problem is
trivial and unique.

In the case of infinitely many variables, however, both issues create problems. As we shall see, and as is intuitively
clear, there is no equivalent in the case of infinitely many zeroes to the notion of degree, so that the requirement that
both functions have the same degree has no meaningful equivalent. 

Similarly, an analogous approach to the determination of $p(z)$ and $q(z)$ from $z_m(0)$ and $\dot{z}_m(0)$
leads to the problem of interpolating 
infinitely many values at infinitely many points by an entire function. This turns out to be a difficult problem, discussed in 
several works \cite{gelfond,groza}. The most obvious issue is that we cannot count count the number of values to be fitted 
to obtain the degree of the fitting polynomial. In that case, neither existence nor uniqueness of the interpolating function 
can be guaranteed, nor can we appeal to uniqueness properties for the solutions of the goldfish equations, since uniqueness
results for infinite systems of ODE's are notoriously non-trivial. 

To specify our meaning, we shall always consider the zeroes $z_m(t)$ of the function $\Phi(z,t)=p(z)+tq(z)$
for {\em given\/} entire functions $p(z)$ and  $q(z)$. The question at hand is then whether the $z_m(t)$
satisfy some form of (\ref{eq:1bis}). 

Next we turn to the regularity properties of $z_m(t)$ as a function of $t$. Since the $z_m(t)$ are solutions of the 
equations $\Phi(z,t)=0$, their existence and smoothness properties are determined by the implicit function theorem. 
For it to hold, we require that the derivative with respect to $z$ of $\Phi(z,t)$ at $z_m(t)$ does not vanish, in other words, that
$z_m(t)$ should not be a {\em double zero\/} of $\Phi(z,t)$. The double zeroes of $\Phi(z,t)$ correspond to isolated values 
of both $z$ and $t$, and are thus avoided by a generic choice of path for the variable $t$. Following this line of argument,
it is readily seen that the $z_m(t)$ are analytic in $t$ in a neighborhood of such a path $C$ for $t$. 
Such proofs are straightforward, since the distance from any point of a given curve $C$ to the nearest double zero 
is always bounded from below, so that the implicit function theorem yields, around any value of $t$, an interval of 
minimal prescribed length in which the equation has a solution.

Again, entire functions behave in a significantly different way, since they can have sequences of pairs of zeroes, the distance 
of which tend to zero, thereby making the application of the implicit function theorem problematic. 

%

\section{Entire functions}

The theory of entire functions is a highly developed and rather intricate part of mathematics. In order not to get
involved in excessive technicalities, we shall limit ourselves to entire functions of order $\rho<1$. 
For greater detail, the reader may consult \cite{entire1,entire2}.
Functions of order $\rho$
are defined by the conditions:
\begin{subequations}
\begin{eqnarray}
&&\limsup_{r\to\infty}\frac{\ln\ln M(r)}{\ln r}=\rho<\infty
\label{eq:5a}
\\
&&M(r)=\max_{|z|=r}|f(z)|,
\label{eq:5b}
\end{eqnarray}
\label{eq:5}
\end{subequations}
which is a precise formulation of the notion that $f(z)$ grows ``roughly'' as $\exp(|z|^\rho)$ as $|z|\to\infty$.
If the $z_n$ are the zeroes of a function $f(z)$ of order $\rho$ such that $f(0)=1$, then it follows from results by Hadamard, that $f(z)$
has the following unique representation
\begin{equation}
f(z)=\prod_{m=1}^\infty \left(
1-\frac{z}{z_m}
\right)
\label{eq:prodinf}
\end{equation}
In other words, the representation is identical to that we just studied for polynomials. Furthermore, again due to results of Hadamard,
it can be shown that a function has order $\rho$ if and only if
\begin{equation}
\sum_{n=1}^\infty\left|
z_n
\right|^{-\rho-\epsilon}<\infty,
\label{eq:7}
\end{equation}
for all $\epsilon>0$,
at least for $\rho$ not an integer, which will always be the case, since we always have $\rho<1$. This therefore
suggests that we should be able to carry over the techniques used for the case of polynomials to
entire functions of order $\rho<1$. {   Parenthetically note that for $\rho\geq1$ 
the representation is significantly more complex, 
since in that case the product (\ref{eq:prodinf}) diverges. The resulting formula involves an exponential prefactor
and a modified form of the product. The analogy with polynomials is then much less clear.} 
\subsection{A singular example}
\label{subsec:irregular}

\begin{figure}
\begin{center}
\includegraphics[scale=0.9]{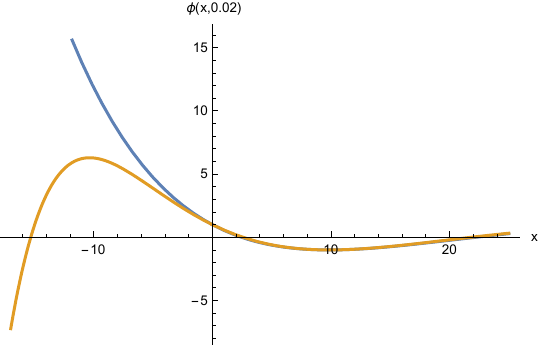}
\end{center}
\caption{
Plot of the function $\cos(\sqrt x)+0.02[1-\cos(2\sqrt x)]$ and of the unperturbed
function $\cos(\sqrt x)$. The positive zeroes of both functions are indeed close, 
whereas the  large negative zero is wholly unrelated
to $\cos(\sqrt x)$, which for $x<0$, is equal to $\cosh(\sqrt{|x|})$ and thus has no zeroes. 
}
\label{fig:1}
\end{figure}

Here we give a concrete example showing that the behavior described above for $p(z)$ and $q(z)$ polynomials 
of different degrees, occurs for entire functions of the same order, so that the order of an entire function
provides no substitute for the degree: let $z(t)$ be the root of
\begin{equation}
\Phi(z,t)=\cos\left(
\sqrt z
\right)+
t\left[
1-\cos\left(2
\sqrt z
\right)
\right]
\label{eq:8}
\end{equation}
$\Phi(z,t)$ is a linear combination of entire functions of order $1/2$.
Clearly the $z_m(0)$ are the zeroes of $\cos(\sqrt z)$, hence $\pi^2(m+1/2)^2$, whereas $z_m(t)$ 
can be evaluated exactly, and has two branches
\begin{equation}
z_{m,\pm}(t)=\arccos^2(w_\pm),\qquad w_\pm=\frac{-1\pm\sqrt{1+16t^2}}{4t}.
\label{eq:9}
\end{equation}
Since $|w_+|<1$ for all real $t$, it is clear that the corresponding value of $z_{m,+}(t)$ is real, and connects to the 
initial value $z_m(0)$. On the other hand, the $w_-$ diverges as $t\to0$ as $-1/(2t)$. For $t\ll1$ one has a real
zero of the form $\arccos^2(w_-)=-\mbox{\rm arccosh}^2(|w_-|)<0$, which diverges as  $-\ln(t)^2$. Other 
(complex) zeroes arise from the other branches of the $\arccos$, but also diverge as $t\to0$. For this reason, these 
zeroes have no connection whatever to the initial values $z_m(0)$. They are not, however, false solutions
introduced by the computation: they really fulfil $\Phi(z_{m,-}(t),t)=0$, as shown in Figure \ref{fig:1}.

This shows that $\Phi(z,t)$ here behaves as if the two functions $\cos(\sqrt z)$ and $\cos(2\sqrt z)$ were of 
``different degree''. Of course, there is no question of the number of zeroes being different, but it is of course
true that the {\em density of zeroes}---appropriately defined---is twice as large in the second function 
as in the first.

In this case, the finer classification of entire functions according to {\em type\/} shows 
that $\cos(2\sqrt z)$ has higher type
than $\cos(\sqrt z)$, where the type $\sigma$ of a function of order $\rho$ is defined by
\begin{equation}
\sigma:=\limsup_{r\to\infty}\frac{\ln M(r)}{r^\rho}
\end{equation}
In this case, the type of $\cos(2\sqrt z)$ is 2, whereas that of $\cos\sqrt z$ is 1. However, even the 
classification using types is not sufficient to make sure that there are no zeroes appearing discontinuously
in the way described above. 

%
%

\subsection{Regular solutions of the goldfish equations}

To avoid the difficulties described above, we define the problem in such a way that they 
are assumed not to occur. In other words, 
we choose $p(z)$ and $q(z)$ normalized as always, by $p(0)=1$ and $q(0)=0$, so that 
\begin{equation}
\Phi(z,t)=p(z)+tq(z)=:\prod_{k=1}^\infty\left(
1-\frac{z}{z_k(t)}
\right)
\label{eq:def-zeros}
\end{equation}
where the $z_k(t)$ are {\em defined\/} by the last equality, and {\em assumed\/} to be continuous 
throughout the interval $[0,T]$, {    that is, it is required that the pathologies described in Section 
\ref{subsec:irregular}  do not arise. Further, we need to see to take care that no double zeroes
exist at any times, since the equations (\ref{eq:1bis}) are then ill-defined. I will show that 
double zeroes of $\Phi(z,t)$ only occur for a {\em denumerable\/} set of points $(t_\alpha,z_\alpha)$.
A simple consideration of cardinality then shows that there exists a real number $\phi$
such that the straight line $C$ defined by the relation $z=e^{i\phi} t$ avoids all double zeroes
of $\Phi(z,t)$. We shall always evaluate the solution of (\ref{eq:1bis}) with $t$ taking values
along $C$, thereby avoiding double zeroes.We shall also frequently speak as if $t$ took real values,
something which can always be assumed after an appropriate straightforward redefiniton of $p(z)$
and $q(z)$.}

To show {   that the double zeroes of $\Phi(z,t)$ are denumerable}, we first define $p_N(z)$ (respectively $q_N(z)$) as the 
polynomial having the same first $N$ zeroes 
as $p(z)$ (respectively $q(z)$) and the same normalization.  Then define $\Phi_N(z.t)$ as $p_N(z)+tq_N(z)$,
and $z_k^{(N)}(t)$ are defined to be the zeroes of $\Phi_N(z,t)$. It is, of course, by no means true
that $\Phi_N(z,t)$ has the same first $N$ zeroes as $\Phi(z,t)$. {   From elementary results on polynomials
it follows that the double zeroes of $\Phi_N(z,t)$ are finite in number, given by $(z_\alpha^{(N)},t_\alpha^{(N)})$ 
(this follows from the fact that double zeroes are places where the discriminant $\Delta(t)$
of $\Phi_N(z,t)$  vanishes.
Since $\Delta(t)$ is a polynomial, the result follows, except if $\Delta(t)$ vanishes identically in $t$).
Since the $\Phi_N(z,t)$ converge uniformly on compact sets towards $\Phi(z,t)$, the zeroes of the
$\Phi_N(z,t)$ converge to those of $\Phi(z,t)$, and hence the double zeroes of the $\Phi_N(z,t)$  tend to those
of $\Phi(z,t)$, showing that, as stated above, the double zeroes of $\Phi(z,t)$ are denumerable.
Whether the $z_\alpha$ are isolated is not clear to me. On the other hand
the $t_\alpha$ need in no way be isolated, and might even be dense. In any case, we shall require neither in the 
following.}

%
Since the $z_m^{(N)}(t)$ are roots of polynomials, these simply obey {   the modified goldfish equations} 
(\ref{eq:1bis}) as they stand, for a finite value of $N$. 

The infinite products defining $p(z)$, $q(z)$ and $\Phi(z,t)$ all converge, and do so uniformly if 
$z$ is limited to a compact set: it is thus clear that, for fixed $k$, $z_k^{(N)}(t)\to z_k(t)$ as $N\to\infty$.
This implies that eventually all $z_k^{(N)}(t)$ converge to the $z_k(t)$ as $N\to\infty$. This however does 
not prove the result we want, since the second derivative of $z_k(t)$ involves an infinite sum of $z_l(t)$, 
which might be poorly approximated by their corresponding finite approximants.

To settle this point we employ the analyticity properties of the $z_m^{(N)}(t)$ {    assumed above}, as well as the 
straightforward consequence 
of Cauchy's theorem, that if a sequence of analytic functions $f_N(z)$ converges uniformly on 
an open set to a function
$f(z)$ as $N\to\infty$, it follows that $f_N^\prime(z)$ similarly converges to $f^\prime (z)$ 
as $N\to\infty$ \cite{ahlfors}. Indeed, the above 
arguments showing the convergence of $z_m^{(N)}(t)$ to $z_m(t)$ do not depend on $t$, as 
long as $t$ remains bounded. 
We thus may assert that $z_m^{(N)}(t)$ tends to $z_m(t)$ uniformly on a fixed neighborhood 
of the real line, and hence that 
$\dot{z}_m^{(N)}(t)$ tends to $\dot{z}_m(t)$ in a similar manner. 

{   We now need to show that the limit of the $z_m^{(N)}(t)$ in fact satisfy (\ref{eq:1bis}). To this
end,} we take the derivative with respect to $t$ of the equation defining $\Phi(z,t)$ and $\Phi_N(z,t)$, 
yielding
\begin{subequations}
\begin{eqnarray}
q(z)&=&z\sum_{k=1}^\infty\frac{\dot{z}_k(t)}{[z_k(t)]^2}\prod_{m=1;(m\neq k)}^\infty\left(
1-\frac{z}{z_m(t)}
\right)\\
q_N(z)&=&z\sum_{k=1}^N\frac{\dot{z}_k^{(N)}(t)}{[z_k^{(N)}(t)]^2}\prod_{m=1;(m\neq k)}^\infty\left(
1-\frac{z}{z_m^{(N)}(t)}
\right)
\end{eqnarray}
\end{subequations}
from which follows
\begin{subequations}
\begin{eqnarray}
q[z_m(t)]&=&\frac{\dot{z}_m(t)}{z_m(t)}\prod_{l=1;(l\neq m)}^\infty\left(
1-\frac{z_m(t)}{z_l(t)}
\right)
\label{eq:wowa}
\\
q_N[z_m^{(N)}(t)]&=&\frac{\dot{z}_m^{(N)}(t)}{z_m^{(N)}(t)}\prod_{l=1;(l\neq m)}^N\left(
1-\frac{z_m^{(N)}(t)}{z_l^{(N)}(t)}
\right)
\label{eq:wowb}
\end{eqnarray}
\label{eq:wow}
\end{subequations}
Since the left-hand side of (\ref{eq:wowb}) tends to that of (\ref{eq:wowa}) and since the prefactor of the 
right-hand side does so as well, we are left with
\begin{equation}
\prod_{l=1;(l\neq m)}^N\left(
1-\frac{z_m^{(N)}(t)}{z_l^{(N)}(t)}
\right)
\to\prod_{l=1;(l\neq m)}^\infty\left(
1-\frac{z_m(t)}{z_l(t)}
\right),
\label{eq:final}
\end{equation}
{    as $N\to\infty$. A direct analysis of this product would be delicate, since it involves values of $z_l^{(N)}(t)$ with $l$ 
comparable to $N$, for which the fixed-$l$ convergence established above provides no information; the argument 
above avoids this difficulty by extracting the limit indirectly, through division of the two quantities already shown to converge.}

From the convergence stated in (\ref{eq:final}) we may, due to analyticity on a common domain, take the derivative
on both sides of the convergence statement, which leads to the statement that the $z_m(t)$ satisfy the system of ODE's  
(\ref{eq:1bis}) with $N$ equal to infinity.

\subsection{A condition implying the hypotheses}

We have drawn the interesting conclusion that the zeroes of $\Phi(z,t)$ obey the infinite set of goldfish equations 
(\ref{eq:1bis}) under the assumption that these solutions satisfy an appropriate regularity condition. and that the
$z_m(t)$ are analytic on a common domain. 

Here we provide 
a condition sufficient to guarantee {   both the regularity and the analyticity assumptions.
This condition is} similar to that of the degree of $q(z)$ being less than that of $p(z)$ when 
both are polynomials. In the latter case,
it is clear that 
\begin{equation}
\lim_{|z|\to\infty}\left|\frac{q(z)}{p(z)}\right|=0
\label{eq:poldeg}
\end{equation}
For entire functions, we must consider the fact that these generally
vanish infinitely often, so we modify this condition to: there exists a sequence of $R_j$ with $R_j\to\infty$
as $j\to\infty$ such that
\begin{equation}
\lim_{j\to\infty}\,\max_{|z|=R_j}\left|\frac{q(z)}{p(z)}\right|=0.
\label{eq:fundeg}
\end{equation}
We denote this relation by $q(z)\ll p(z)$. It is, for instance, implied if the order of $q(z)$ is less than that of $p(z)$, 
or if, their orders being equal, the type of $q(z)$ is less than that of $p(z)$. The condition is quite weak, so that the 
result which has this as a hypothesis is quite strong. 

However, it is different from the polynomial case in an essential manner: the degree varies discontinuously, so that 
two polynomials can satisfy $p_1(z)\ll p_2(z)$ and yet there exists no polynomial $p_3(z)$ such that $p_1(z)\ll p_3(z)$
and $p_3(z)\ll p_2(z)$, but the opposite property obviously holds for non-polynomial entire functions.
 {   Further, the relation $p_1(z)\ll p_2(z)$ by no means {    defines even a partial order:} on the one hand,
 it is possible that $p_1(z)\ll p_2(z)$ and $p_2(z)\ll p_1(z)$ both hold, on the other, the relation is not transitive: it is possible 
to find functions $p_1(z)$, $p_2(z)$ and $p_3(z)$ such that $p_2$ dominates $p_1$ on a certain set of radii growing to infinity, 
$p_3$ dominates $p_2$ on another such set, whereas $p_3$ and $p_1$ do not dominate each other on any sequence of radii 
\cite{ref}}.

To prove the result we use Rouch\'e's theorem \cite{ahlfors}: 
if $f(z)$ and $g(z)$ are analytic inside the closed contour $C$ and $|g(z)|<|f(z)|$ on $C$, then $f(z)$ and $f(z)+g(z)$
have the same number of zeroes inside $C$. From $q(z)\ll p(z)$, it follows that the hypothesis of Rouch\'e's theorem
holds for any $t$ if we take $f(z)$ to be $p(z)$, $g(z)$ to be $tq(z)$, and $R_j$ to be large enough. Thus 
inside a circle of radius $R_j$ sufficiently large, $p(z)$ and $\Phi(z,t)$ have the same number of zeroes
for all $t\leq T$ where $T$ can be chosen arbitrary but fixed. From this evidently follows the claim that the $z_m(t)$
are continuous for $0\leq t \leq T$, and since $T$ is arbitrary, for all $t$. This is quite similar, both in its statement and
its proof, to a theorem by Hurwitz \cite{hurwitz} stating, in rough terms, that if the analytic functions $f_n(z)$ tend to 
$f(z)$, then the zeroes of $f_n(z)$ tend to those of $f(z)$. 

Let us now discuss the analyticity of the $z_m(t)$ with respect to $t$. To this end we allow $t$ 
freely to vary over the entire complex plane. Since the $z_m(t)$ are solutions of the equations 
$\Phi(z,t)=0$, the implicit function
theorem guarantees locally the existence of such solutions, and the analyticity of $\Phi( z,t)$ both in $z$ and 
in $t$ implies that of the $z_m(t)$. However, the implicit function theorem requires the
nonvanishing of the derivative of $\Phi(z,t)$ with respect to $z$. That means that the application of
the theorem fails at such values of $t$ for which the function $\Phi(z,t)$
has a double zero {    at $z_m(t)$, that is, whenever}.
\begin{equation}
\prod_{k=1(\neq m)}^\infty\left[
z_m(t)-z_k(t)
\right]=0
\end{equation}
{    This means that the function $z_m(t)$ is analytic in $t$ within a circle of radius equal to the distance  
to the complex time 
$t_{coll}$ such that the zero $z_m(t_{coll})$ collides with some other zero $z_l(t_{coll})$. Note in particular that collisions between
two zeroes both different from $z_m(t)$ {\em do not affect\/} the analyticity of $z_m(t)$.}

{    From these considerations it follows that the convergence radius of $z_m^{(N)}(t)$ has an 
{\em $N$-independent lower bound\/} over a bounded time interval $[0,T]$. Indeed, for any fixed $N$, only a finite,
$N$-independent number of zeroes can reach a given $z_m(t)$ in a time $t\in[0,T]$. This follows from the 
assumption that $q(z)\ll p(z)$
in the following manner: for a given set of functions $z_m^{(N)}(t)$ there is a given radius $R$ depending on $T$ 
but not on $N$ such that zeroes of $\Phi_N(z,0)$ outside of the circle of radius $R$ never cross that circle within time $T$.
To show this, choose a radius $R$ such that $q(z)/p(z)$ is  sufficiently small, which is always possible 
due to the hypothesis $q(z)\ll p(z)$.
Applying Rouch\'e's theorem, we find that 
the number of zeroes of $\Phi(z,t)$ inside the circle of radius of $R$
remains equal to that of $\Phi(z,0)$ for all $t\in[0,T]$.Thus no zero outside the circle of radius $R$ can cross it in this time,
and hence the function $z_m(t)$ can only collide with those zeroes inside the 
circle of radius $R$, which are finite in number, since an infinite number of zeroes in a compact domain 
would contradict the property of being an entire function.}

{    Since this radius $R$ depends on $T$ but can be determined for the entire function $\Phi(z,t)$
similar properties hold for the zeroes of the finite approximants $p_N(z)$, $q_N(z)$ and $\Phi_N(z,t)$. Thus we see that
the choice of $R$ is $N$-independent. As a consequence the zero $z_m^{(N)}(t)$ of $\Phi_N(z,t)$ 
can only collide with a finite number of the other zeroes of $\Phi_N(z,t)$ within a time $T$. Since 
the curve $C$ on which $t$ moves never hits a double zero, the convergence radius remains bounded from below 
by a non-zero $N$-independent bound, and $z_m^{(N)}(t)$ can thus be analytically continued over the entire 
interval of $C$ corresponding to $0\leq t\leq T$. The hypothesis that all $z_m^{(N)}(t)$ have a common
$N$-independent domain of analyticity is follows therefore from the hypothesis that $q(z)\ll p(z)$. From this also 
straightforwardly follows that $z_m(t)$ is analytic on that same domain. 
}

\section{Conclusions}

Let us see what has been found and what remains open. We have shown that, if $p(z)$ and $q(z)$ are two entire functions
of order $\rho<1$ satisfying $q(z)\ll p(z)$, then the zeroes of $p(z)+tq(z)$ satisfy the system of ODE's (\ref{eq:1bis})
with $N$ set to infinity and with the initial conditions arising from the conditions:
\begin{subequations}
\begin{eqnarray}
p[z_m(0)]&=&0
\label{eq:condinita}
\\
\dot{z}_m(0)&=&q[z_m(0)]z_m(0)
\prod_{l=1;(l\neq m)}^\infty
\left(
1-\frac{z_m(0)}{z_l(0)}
\right)^{-1}
\label{eq:condinitb}
\end{eqnarray}
\label{eq:condinit}
\end{subequations}
What is missing is the converse: there is no guarantee that all initial conditions can be generated in this manner,
nor is it clear that how to describe the set of initial conditions for which a solution of the infinite systems (\ref{eq:1bis}) 
exists. Thus we have not described the dynamics generated by this system, indeed we have not characterized this dynamics 
at all, nor shown that it exists {   except through the indirect procedure outlined in this paper}.

{   
The difficulty in this problem is that $q(z)$ is determined by an infinite interpolation problem, namely its values
on the $z_m(0)$ are known. However, the condition $q(z)\ll p(z)$ provides no information on whether the problem has a 
solution. The problem arises from the fact that the growth of the norms of the zeroes can be extremely irregular if it is only
limited by the condition of the function having order $\rho$ or less. Imposing  more stringent conditions 
might be of interest, such as entire functions of completely regular growth, as described in \cite{entire1}, but this is reserved for future work. 
}

%

Another non-trivial issue is the extent to which this can be viewed as a solution. As matters stand, we have proved a connection 
between a poorly characterized infinite dynamics and the equally opaque null set of a known entire function. Indeed, null sets
of polynomials are well-understood objects, so that connecting them to the solutions of a system of nonlinear ODE's
counts as genuine progress. A similar feat with polynomials replaced by entire functions is distinctly less impressive.
What, for instance, does the null set of
\begin{equation}
\Phi(z,t)=\cos(\sqrt z)\cos(\alpha\sqrt z)+tz
\end{equation}
look like, if $\alpha$ is, for instance, a Liouville irrational? The function $p(z)$ is of order $\rho=1/2$, whereas $q(z)$ 
is a polynomial, so the problem is well-posed. However, the initial null set contains zeroes which, although
isolated, have nearest-neighbor distances tending to zero.

{   My real motivation for starting this 
apparently weird line of work in the first place was, however, the following open problem: 
there is another system that can similarly be exactly solved \cite{goldfish-per,periods}:
\begin{equation}
\ddot z_j=2\dot z_j\sum^N_{k=1}{}^\prime\frac{\dot z_k}{z_j-z_k}+i\omega\dot{z}_j
\label{eq:1per}
\end{equation}
as it transforms to the equation
\begin{equation}
\frac{d^2}{dt^2}\Phi\left[
z(t),t
\right]+i\omega \frac{d}{dt}\Phi\left[
z(t),t
\right]=0.
\label{eq:4}
\end{equation}
For real values of $\omega$, all solutions of (\ref{eq:4}) are periodic with period $T=2\pi/\omega$.
However, if the polynomial $\Phi$ returns to its original value, this does not mean that the
$z_j(t)$ do so, since they are only determined by $\Phi$ up to order. After a period $T$, in general, the
$z_j(T)$ correspond to a given permutation of the $z_j(0)$, so that the period of the
$z_j(t)$ is of the form $mT$ with $m$ an integer typically far smaller than $N!$ \cite{goldfish-per,periods}.}

In such a system, the function $\Phi(z,t)$ varies periodically in time with period $T=2\pi/\omega$. Since the
zeroes $z_m(t)$ are only determined up to order, however, their motion is generally more intricate: 
in the polynomial case, the only issue is that the zeroes return after a period 
affected by a permutation that depends on the initial condition. Since the zeroes are in finite number,
the permutation will eventually yield the identity after
sufficiently many repetitions, so that the zeroes' motion is periodic with a primitive period $mT$, 
where $m$ is an integer depending on the initial condition \cite{periods}.

When extended to infinitely many particles, though, the possibilities become, depending on one's viewpoint, 
alarming or fascinating. It is at least within the realm of possibility that the $z_m(t)$ in such a case might behave 
in a highly complex, even perhaps a chaotic manner.

Francesco Calogero had for long been searching for systems which might combine chaos with 
integrability \cite{riemann1, riemann2}. I believe the possibilities opened by these speculations might 
have afforded him pleasure, and dedicate these thoughts to his memory.

\end{document}